\documentclass[12pt,a4paper,twocolumn,fleqn]{narms}

\usepackage{subcaption}
\usepackage{epsfig}
\usepackage{timesmt}

\usepackage[utf8]{inputenc}
\usepackage[T1]{fontenc}
\usepackage[english]{babel}
\usepackage{amsmath}
\usepackage{amsfonts}
\usepackage{color,soul}
\usepackage{amssymb}
\usepackage{graphicx}
\usepackage{siunitx}
\usepackage{url}
\usepackage{float}

\usepackage{chicaco}

\begin{document}
\title{Bayesian neural networks for the probabilistic forecasting of wind direction and speed using ocean data}
\author{{Mariana C. A. Clare \& Matthew D. Piggott} \\
{\aff{Department of Earth Science and Engineering}} \\
{\aff{Imperial College London, London, UK.}}
}

\date{}

\date{}

\abstract{Neural networks are increasingly being used in a variety of settings to predict wind direction and speed, two of the most important factors for estimating the potential power output of a wind farm. However, these predictions are arguably of limited value because classical neural networks lack the ability to express uncertainty. Here we instead consider the use of Bayesian Neural Networks (BNNs), for which the weights, biases and outputs are distributions rather than deterministic point values. This allows for the evaluation of both epistemic and aleatoric uncertainty and leads to well-calibrated uncertainty predictions of both wind speed and power. 
    Here we consider the application of BNNs to the problem of offshore wind resource prediction for renewable energy applications.
    For our dataset, we use observations recorded at the FINO1 research platform in the North Sea and our predictors are ocean data such as water temperature and current direction. 
    
    \hspace{5pt} The probabilistic forecast predicted by the BNN adds considerable value to the results and, in particular, informs the user of the network's ability to make predictions of out-of-sample datapoints. We use this property of BNNs to conclude that the accuracy and uncertainty of the wind speed and direction predictions made by our network are unaffected by the construction of the nearby Alpha Ventus wind farm. Hence, at this site, networks trained on pre-farm ocean data can be used to accurately predict wind field information from ocean data after the wind farm has been constructed.}

\maketitle

\section{Introduction}\label{sec:intro}
 Two of the most important factors required to predict and optimise the power generated by a wind farm are wind speed and wind direction. Wind speed is particularly difficult to estimate because of the intermittent nature of wind \cite{khosravi2018prediction}. Traditionally a range of physical, statistical and hybrid methods have been used to approximate the wind speed, but recently neural networks have begun to be used \cite{brahimi2019using}. This is in line with the increasing use of AI techniques to optimise and control wind generation in wind farms \cite{wang2020review}. Neural network predictions of wind speed, direction and power are usually tailored to provide short to medium-term forecasts \cite{rotich2014forecasting,kim2018short,khosravi2018time,zucatelli2019short}, but there has been some research into predicting wind speed, direction and power from other met-ocean variables \cite{antorcomparison,brahimi2019using,khosravi2018prediction}, which is of more value for long-term optimisation of wind farm design. \citeN{Baumgartner2020} even use climate re-analysis data to predict wind power output for the whole of Germany for a 7-year time series, although we note that in their study, wind speed is an input variable that is considered known. 
 
Classical neural network predictions produce deterministic estimates, which make it difficult to assess their uncertainty and limits the extent to which they can be useful. A possible solution to this issue is use an ensemble approach (e.g. \citeNP{kim2018short}), but choosing a good ensemble of models is non-trivial \cite{Scher2020} and may be computationally expensive because it requires the network to be trained multiple times. A different approach is to follow \citeN{Gal2016} and use Monte Carlo dropout to create a Bayesian approximation. This technique is used in \citeN{Karami2021} to assess uncertainty in the relation between wind speed and wind power output for specific wind turbines. However, this only assesses epistemic uncertainty (uncertainty in the model) rather than aleatoric uncertainty (uncertainty in input data). 

A more sophisticated technique is to use a Bayesian Neural Network (BNN) where the weights, biases and outputs are distributions rather than deterministic point values, allowing for the evaluation of both epistemic and aleatoric uncertainty and leading to well-calibrated uncertainty predictions \cite{jospin2020hands}. BNNs have already been successfully applied to short term forecasting of wind power in \citeN{Mbuvha2017}. Their ability to predict the uncertainty of the outcome is particularly important for future wind farm predictions because there is a substantial potential for out-of-sample datapoints. This is both due to the fact that the presence of wind farms may alter met-ocean data, and that a changing climate may mean that future data is no longer consistent with historical data. In this work, we apply BNNs to weather data to predict wind direction and wind speed in an offshore setting. We train the BNN using met-ocean data from the FINO1 research platform located in the North Sea \cite{fino1,fino_data}. In 2010, the Alpha Ventus wind form was built in close proximity to the research platform (the closest turbine is approximately \SI{400}{\metre} away). Studies of both the FINO1 data and other data in the area have seen clear evidence of changes in met-ocean flow conditions due to the presence of this wind farm \cite{platis2018first,ortensi2020long,Barfuss2021}. Therefore some works separate the FINO1 dataset into pre- and post-construction phases (e.g. \citeNP{Pettas2021}). However, we choose to train and test the BNN on the pre-wind-farm data and then also test it on post-wind-farm data. This allows us to investigate whether our BNN is capable of making accurate predictions with low uncertainty after the environment change. Finally, we note that, as we are considering both meteorological and ocean variables (\textit{i.e.}, met-ocean variables), the analysis in this work relates only to offshore wind farms and not to onshore ones.

The remainder of this work is structured as follows: Section \ref{sec:data} explores the FINO1 dataset and selects the input variables for training; Section \ref{sec:nn_arch} describes and tunes the neural network architecture; Section \ref{sec:results} discusses the results of the BNN; and Section \ref{sec:conc} concludes the work.

\section{Data Selection}\label{sec:data}
In this work, we use data from the FINO1 research platform available at \citeN{fino_data}. This dataset consists of both meteorological and oceanographic variables spanning from 2004 to 2021, although there are large periods of missing data and not all variables have been collected since the inception of the research platform. The timestamps of the data are not always consistent across variables and the frequency of the raw data varies between 10 and 30 minutes depending on the variable. To manage this we take the average of each variable over an hourly period. Our target variables are the wind direction and wind speed at \SI{91}{\metre} above the lowest astronomical tide (LAT), which is roughly the hub height of the turbines at the Alpha Ventus wind farm installed in close proximity to the FINO1 research platform \cite{Pettas2021}. To determine which input variables to use for the wind speed predictions and wind direction predictions, we use the `ExtraTreesRegressor' model in scikit-learn for feature importance analysis \cite{scikit-learn}. In our preliminary dataset, we include water temperature and current direction (at various ocean heights), average wave period, significant wave height and the timestamp of the data split into year, month, day and hour. This gives a total of 25 features and Table \ref{table_feature_dir} and Table \ref{table_feature_speed} show the top 12 most important features for predicting the wind speed and direction at \SI{91}{\metre}, respectively, according to the feature importance analysis. These 12 features form the inputs of each prediction dataset. Notably they are all oceanographic variables which can be measured without the need to build an expensive tall research platform.

For our training dataset, we use the years 2005 and 2006 (corresponding to over 13,000 rows of data), for validation we use the shuffle mechanism, and for our test dataset we use the year 2004 (corresponding to over 3000 rows of data). We emphasise that both the training and test dataset are pre-2010, \textit{i.e.} before the Alpha Ventus wind farm was built. This is so that we can first test whether a neural network trained on a dataset before a wind farm is built is capable of making accurate predictions before the wind farm changes met-ocean conditions. We also consider a second dataset from 2010 to 2021 after the wind farm was built (corresponding to over 34,000 rows of data). Recall that for this data, the met-ocean conditions have changed due to the presence of a wind farm \cite{Pettas2021}.

\begin{table}[H]
	\centering
	\begin{tabular}{c|c}
\textbf{Feature} & \textbf{\begin{tabular}[c]{@{}c@{}}Feature \\ Importance\end{tabular}} \\ \hline
		Current direction (2m)                                           & 0.198                                           \\
		Current direction (0m)                                           & 0.136                                           \\
		Current direction (4m)                                           & 0.126                                           \\
		Year                                                          & 0.0742                                          \\
		Significant waveheight                                        & 0.0449                                          \\ 
		Average wave-period                                           & 0.0416                                          \\
		Day                                                           & 0.0340                                          \\
		Current direction (6m)                                           & 0.0339                                          \\
		Month                                                         & 0.0239                                          \\ 
		Hour                                                          & 0.0237                                          \\
		Current direction (16m)                                          & 0.0192                                          \\
		Current direction (28m)                                          & 0.0186                                         
	\end{tabular}
\caption{Top 12 most important features for predicting wind direction, where current direction height values given in brackets are the depths below LAT.}\label{table_feature_dir}
\end{table}

\begin{table}[H]
		\centering
	\begin{tabular}{c|c}
		\textbf{Feature} & \textbf{\begin{tabular}[c]{@{}c@{}}Feature \\ Importance\end{tabular}} \\ \hline
	Significant waveheight  &	0.429444 \\
	Average wave-period &	0.108533 \\
	Current direction (0m) &	0.090767 \\
	Water temperature (6m)	&  0.044099 \\
	Month	& 0.036995 \\
	Day	& 0.031185 \\
	Year &	0.031057\\
	Water temperature (25m) &	0.029656 \\
	Water temperature (3m)	& 0.022675 \\
	Water temperature (20m) &	0.020927 \\
	Hour &	0.019324
	\end{tabular}
\caption{Top 12 most important features for predicting wind speed, where current direction and water temperature heights are the heights below lowest astronomical tide (LAT).}\label{table_feature_speed}
\end{table}

\section{Neural network architecture and tuning}\label{sec:nn_arch}
\begin{figure}
	\begin{subfigure}{0.47\textwidth}
		\centering
		\includegraphics[width=0.9\textwidth]{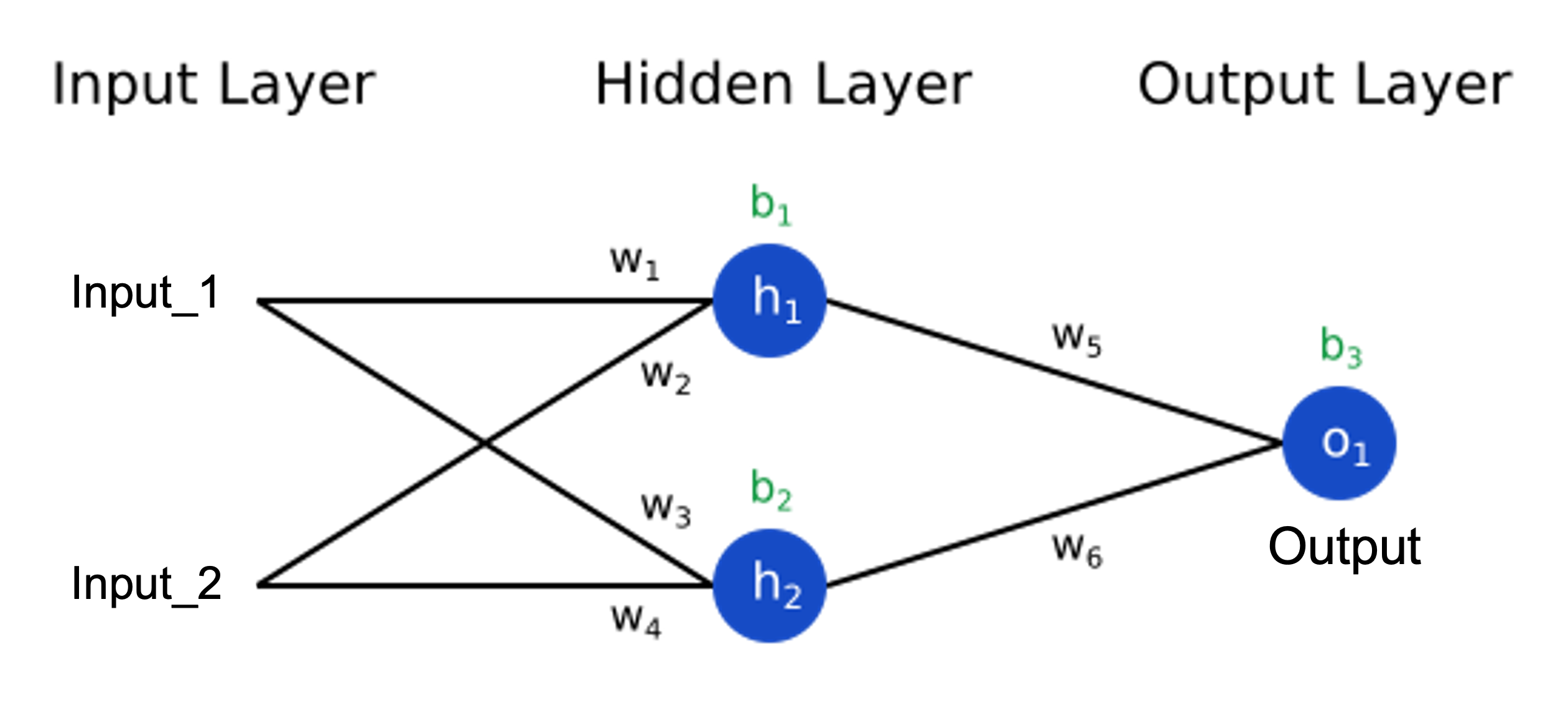}
		\caption{Classical Neural Network. Weights and biases are point estimates.}
		\label{fig:classical_nn}
	\end{subfigure}
	\hfill
	\begin{subfigure}{0.47\textwidth}
		\centering
		\includegraphics[width=0.9\textwidth]{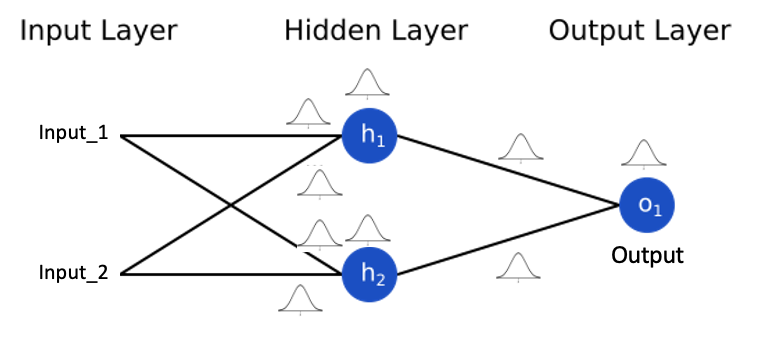}
		\caption{Bayesian Neural Network (BNN). Weights and biases are distributions.}
		\label{fig:BNN}
	\end{subfigure}
	\caption{Comparing a standard neural network to a BNN.}\label{fig:nn}
\end{figure}

As discussed in Section \ref{sec:intro}, BNNs differ from classical neural networks because the weight and biases on at least some of the layers are distributions rather than single point estimates (see Figure \ref{fig:nn}). More specifically, as BNNs use a Bayesian framework, once trained these distributions are the posterior distributions based on the training data \cite{Bykov2020}. Note that for brevity in this section, we refer to the weights and biases as network parameters. The distributions in the output layer represent aleatoric uncertainty (uncertainty in the data) and the distributions in the hidden layers represent epistemic uncertainty (uncertainty in the model) \cite{salama_2021}. The latter can be reduced by increasing the size of the training dataset and therefore can also be seen as a measure of uncertainty due to insufficient data. 

Following \citeN{jospin2020hands}, the posterior distributions in the BNNs are calculated following Bayes rule
\begin{equation}
\begin{split}\label{eq:bates}
p\left(W\lvert D_{tr}\right) & = \frac{p(D_{tr}\lvert W)p(W)}{p({D_{tr})}}, \\ & = \frac{p(D_{tr}\lvert W)p(W)}{\int_{W} p(D_{tr}\lvert W)p(W)\,dW},
\end{split}
\end{equation}
meaning the probability of output $y$ given input $x$ is then
\begin{equation}
p(y\lvert x, D_{tr}) = \int_{W} p(y\rvert f(x; W))p(W\rvert D_{tr})\, dW.
\end{equation}
Here $W$ are the network parameters, $D_{tr} = \left(x_{n}, y_{n}\right)$ the training data, $p(W)$ the prior distribution of the parameters and $f(\,\cdot\,; W)$ is the neural network. Computing (\ref{eq:bates}) directly is very difficult, especially due to its denominator \cite{jospin2020hands,Bykov2020}. A number of methods have been proposed to approximate this denominator term including Markov Chain Monte Carlo sampling \cite{titterington2004bayesian} and variational inference \cite{osawa2019practical}. We use the latter and refer the reader to \citeN{clare2022explainable}, for example, for more details of how this technique is used with BNNs. We remark that variational inference requires the shape of the posterior distribution to be assumed and in our work, we follow standard practice and assume they are all normal distributions. Furthermore, for all priors in the BNN (used in (\ref{eq:bates}), we use the normal distribution $\mathcal{N}(0, \sigma)$, which is again standard practice because of the normal distribution's mathematical properties and simple log-form \cite{silvestro2020prior}. 

We have chosen to train two separate BNNs in this work: one which predicts wind speed and the other that predicts wind direction. The exact angle of the wind direction is difficult to predict and highly variable over an hourly period. Moreover, the general angle direction is of more value to practitioners and therefore we follow the same idea as in \citeN{clare2021combining} and bin the wind direction data into 8 bins corresponding to [N-NE, NE-E, E-SE, SE-S, S-SW, SW-W, W-NW, NW-N]. Whilst significant waveheight, average wave-period and the date and time are important for both wind direction and wind speed, Table \ref{table_feature_dir} shows that current direction variables are helpful for predicting wind direction, whereas Table \ref{table_feature_speed} shows that water temperature are helpful for predicting wind speed. Current direction is difficult to measure and uncertain and therefore for the wind direction BNN, we choose to assess both the epistemic and aleatoric uncertainty. For the wind speed BNN, we choose to assess only the epistemic uncertainty (\textit{i.e.} only the weights of the hidden layers are distributions), so as to demonstrate the application of two different types of BNN.

In order to find a suitable architecture for these BNNs, we use hyperparameter tuning on each BNN separately. For the tuning we use the keras-tuner library \cite{omalley2019kerastuner} and construct the BNN using the TensorFlow probability library \cite{dillon2017tensorflow}. We note that not all layers in a BNN must be Bayesian and, in fact, \citeN{Brosse2020} find that the best BNN results are achieved when only the final two hidden layers are Bayesian. We therefore set the number of deterministic layers and the number of Bayesian layers as tunable parameters in the parameter search, with the deterministic layers before the Bayesian ones. For both BNNs we find, like \citeN{Brosse2020}, that the optimum number of hidden Bayesian layers is two. We also tune for the number of units for each layer in the network.  For the wind direction BNN, the optimum number of units is [52, 28, 52, 8, 24], where the first three layers are `Dense' and the last two are `DenseVariational' and all use the `tanh' activation function.  The output layer is a [DenseVariational, OneHotCategorical] layer combination from the TensorFlow probability package, which produces a distribution as an output. For the wind speed BNN, the optimum number of units is [24, 24, 16, 16], where the first two layers are `Dense' and the last two are `DenseVariational' and all use the `elu' activation. After each `Dense'/`DenseVariational' layer, there is a `BatchNormalisation' layer and the output layer is then a `Dense' layer with one unit and no activation. Both BNNs are compiled with an Adam Optimizer with an initial learning rate of $10^{-3}$ \cite{kingma2014adam}.

\section{Results}\label{sec:results}
\subsection{Wind direction}
\begin{figure}
	\begin{subfigure}{0.45\textwidth}
		\centering
		\includegraphics[width=0.9\textwidth]{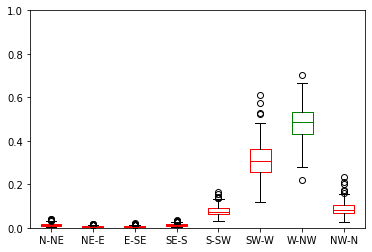}
		\caption{Example where correct bin predicted.}\label{corr_bin}
	\end{subfigure}
	\begin{subfigure}{0.45\textwidth}
		\centering
	\includegraphics[width=0.9\textwidth]{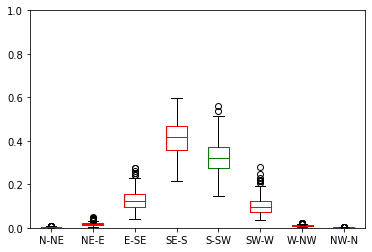}
		\caption{Example where incorrect bin predicted.}
\end{subfigure}
\caption{Box-and-whisker plot of BNN predictions for wind direction, generated using the ensemble members, where the correct bin is coloured in green and the incorrect bins are coloured in red. Note that the points outside the whiskers represent outliers.}\label{bin_examples}
\end{figure}

We first discuss the BNN results for predicting wind direction from ocean data (see Table \ref{table_feature_dir}). Because we are considering aleatoric uncertainty (as well as epistemic uncertainty), the output of our BNN is a distribution. Recall, we can also automatically generate an ensemble of these distributions from the BNN without retraining, due to the distributions in the network parameters. This allows us to generate box plots for these distributions and take into account the epistemic uncertainty of the model. Figure \ref{bin_examples} shows the output at two example datapoints where the ensemble is shown using a box-and-whisker plot. The narrower the box-and-whisker, the more certain the BNN is of the prediction for this bin; for example, in Figure \ref{corr_bin} the BNN is fairly certain that the probability of the wind coming from any of the bins between `NW' and `SW' is low, but there are a range of possible probabilities for the `SW-W' bin and `W-NW' bin. 

\begin{figure}
	\begin{subfigure}{0.45\textwidth}
		\centering
		\includegraphics[width=0.9\textwidth]{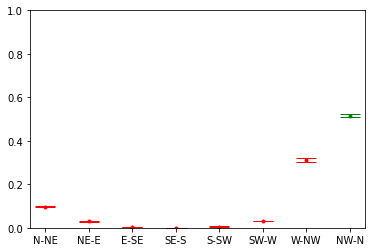}
				\caption{Difference between bin predictions statistically significant.}\label{confidence_interval_dist}
	\end{subfigure}
\hfill
\begin{subfigure}{0.45\textwidth}
\centering
		\includegraphics[width=0.9\textwidth]{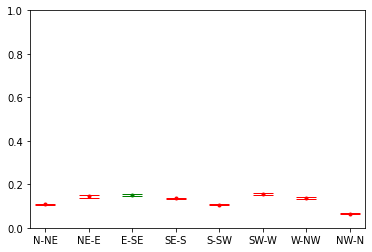}
		\caption{Difference between bin predictions not statistically significant.}\label{confidence_interval_no_dist}
\end{subfigure}
\caption{Confidence interval plot of BNN predictions for wind direction, generated using the ensemble members, where the correct bin is coloured in green and the incorrect bins in red.}
\end{figure}

Both figures shows that there can be significant overlap between the box-and-whisker for each bin. However, this can be misleading as box-and-whisker plots consider upper and lower quartiles which are not relevant for assessing statistical significance. Therefore we also consider confidence intervals to determine whether the differences are statistically significant. Figure \ref{confidence_interval_dist} shows an example of a confidence interval plot where the differences are statistically significant. Focussing only on the post-farm dataset, for 96\% of the datapoints, the probabilities in the ensemble for the most likely bin are statistically different for the probabilities of the other bins. For the remaining datapoints, the probabilities for the top two most likely bins are statistically different from the probability for the other bins. Finally, there are approximately 20 datapoints where even this is not true. An example is shown in \ref{confidence_interval_no_dist}, where almost every bin has the same probability. Although this is not ideal, this is an example of where a BNN has a clear advantage over a classical neural network, because it clearly informs the user that it is very uncertain of its prediction for this datapoint and that using this BNN on this datapoint is inappropriate.

\begin{figure}
	\centering
	\begin{subfigure}{0.45\textwidth}
			\centering
		\includegraphics[width=0.9\textwidth]{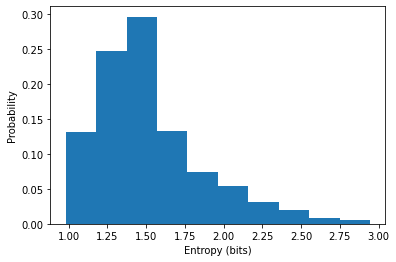}
			\caption{Pre wind farm test dataset.}\label{entropy_pre}
	\end{subfigure}
	\hfill
	\begin{subfigure}{0.45\textwidth}
			\centering
\includegraphics[width=0.9\textwidth]{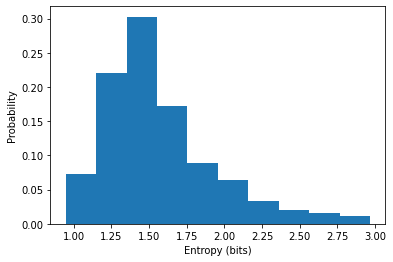}
		\caption{Post wind farm dataset.}\label{entropy_post}
	\end{subfigure}
\caption{Entropy values measuring uncertainty in BNN wind direction predictions before and after the construction of a wind farm}
\end{figure}

To get a numerical measure of the uncertainty expressed by the BNN, we can calculate the entropy of the final distribution. In information theory, entropy is considered as the expected information of a random variable and for each sample $i$ is given by
\begin{equation}
H_{i} = - \sum_{j=1}^{N_{l}} p_{ij}\log(p_{ij}),
\end{equation}
where $N_{l}$ is the number of possible variable outcomes and $p_{ij}$ is the probability of each outcome $j$ for sample $i$ \cite{Goodfellow-et-al-2016}. Hence, the larger the entropy value, the less skewed the distribution and the more uncertain the model is of the result. The entropy plots for pre and post wind farm construction (Figures \ref{entropy_pre}  and  \ref{entropy_post} respectively) are very similar, indicating that the construction of the wind farm has not added uncertainty to our BNN predictions for wind direction. 

\begin{figure}
	\centering
	\includegraphics[width=0.45\textwidth]{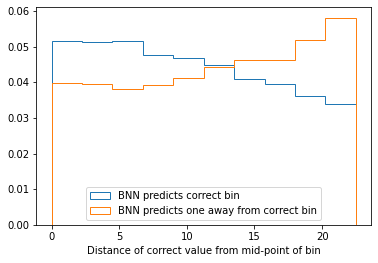}
	\caption{Comparing the distance from the correct value to the mid-point of the bin for the post-farm dataset, when the BNN predicts the correct bin to when the BNN predicts one away from the correct bin.}\label{dist_mid}
\end{figure}

Finally, we discuss the error in our BNN predictions for wind direction. The BNN predicts the correct bin 57\% of the time on the pre-farm test dataset and 55\% of the time on the post-farm dataset. The danger of binning the data means that many errors of bin prediction may be due to the real value being near a bin boundary, resulting in the actual error caused by choosing one bin over another for this datapoint being small. Figure \ref{dist_mid} illustrates this point by showing that when the BNN predicts one away from the correct bin, the correct value tends to be near the bin boundary between the correct bin and the predicted bin. For both the post-farm and pre-farm test datasets, the bin with the second-highest probability is the correct bin for 27\% of the datapoints, and for these cases the second highest probability bin is next to the highest probability bin. Therefore for 84\% of datapoints, the most likely or second most likely bin predicted by the BNN is the correct bin which represents high accuracy for real-world data.

\subsection{Wind speed}
In this work, we have also constructed a BNN for predicting wind speed from ocean data (see Table \ref{table_feature_speed}). As for the wind direction case, when testing our BNN, we generate an ensemble of outputs where each ensemble member is generated simply by using a different sample from the distributions in the weights and biases, with no re-training. We find that the RMSE of the hourly wind speed is $\SI{2.44}{m.s^{-1}}$ and the mean absolute error (MAE) is $\SI{1.96}{m.s^{-1}}$, which is comparable to the errors found using neural neworks to predict wind speed in \cite{antorcomparison}. Moreover, when we apply the BNN to the post wind-farm dataset, we find that the errors on the hourly wind speed are of the same order (RMSE: $\SI{2.63}{m.s^{-1}}$ and MAE: $\SI{2.11}{m.s^{-1}}$). This shows that even though our network is trained on data before the presence of a wind farm, it can still be used to make accurate predictions once the wind farm has been constructed. This is notable because both the ocean conditions \cite{Barfuss2021} and wind conditions \cite{platis2018first} in this area of the North Sea are changed by the presence of the wind turbines. Wind conditions are changed due to turbine wakes, and these wakes cause a reduction in the wave energy due to wind-wave interactions \cite{de2011simulation}. In particular, studies in other areas of the North Sea have shown large reductions in the significant wave height up to \SI{15}{km} from the turbine \cite{christensen2013transmission}, which is one of the key features in our wind speed and direction predictions. Despite this, our BNN is robust and its accuracy and uncertainty is unaffected by these wake changes. The relationship the BNN has learnt from the pre-farm data still holds and can be successfully applied for the post-farm data. Therefore, the BNN can be used to predict conditions for future wind farms.

\begin{figure}
	\centering
\begin{subfigure}{0.2\textwidth}
			\centering
	\includegraphics[width=0.9\textwidth]{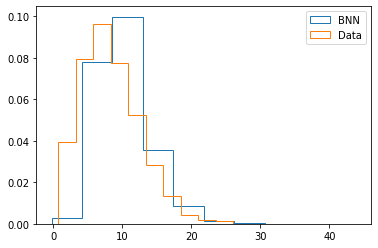}
	\caption{2010}
	\label{fig:2010}
\end{subfigure}
\begin{subfigure}{0.2\textwidth}
	\centering
	\includegraphics[width=0.9\textwidth]{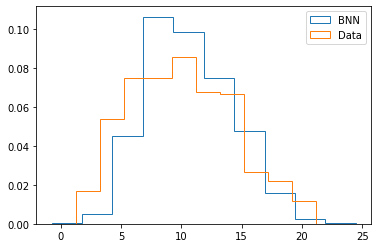}
	\caption{2011}
	\label{fig:2011}
\end{subfigure}
\begin{subfigure}{0.2\textwidth}
	\centering
	\includegraphics[width=0.9\textwidth]{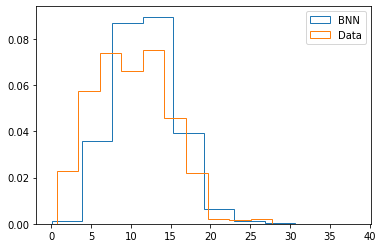}
	\caption{2012}
	\label{fig:2012}
\end{subfigure}
\begin{subfigure}{0.2\textwidth}
	\centering
	\includegraphics[width=0.9\textwidth]{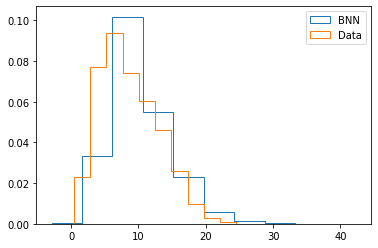}
	\caption{2013}
	\label{fig:2013}
\end{subfigure}
\begin{subfigure}{0.2\textwidth}
	\centering
	\includegraphics[width=0.9\textwidth]{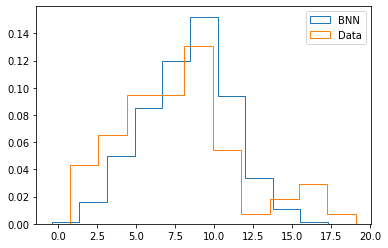}
	\caption{2014}
	\label{fig:2014}
\end{subfigure}
\begin{subfigure}{0.2\textwidth}
	\centering
	\includegraphics[width=0.9\textwidth]{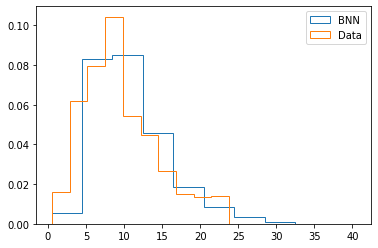}
	\caption{2015}
	\label{fig:2015}
\end{subfigure}
\begin{subfigure}{0.2\textwidth}
	\centering
	\includegraphics[width=0.9\textwidth]{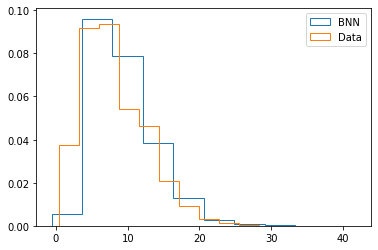}
	\caption{2016}
	\label{fig:2016}
\end{subfigure}
\begin{subfigure}{0.2\textwidth}
	\centering
	\includegraphics[width=0.9\textwidth]{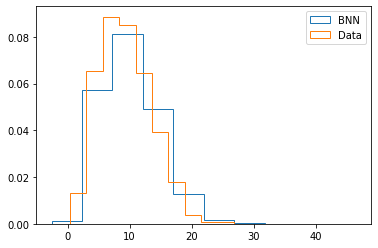}
	\caption{2017}
	\label{fig:2017}
\end{subfigure}
\begin{subfigure}{0.2\textwidth}
	\centering
	\includegraphics[width=0.9\textwidth]{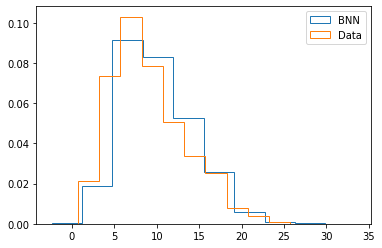}
	\caption{2019}
	\label{fig:2019}
\end{subfigure}
\begin{subfigure}{0.2\textwidth}
	\centering
	\includegraphics[width=0.9\textwidth]{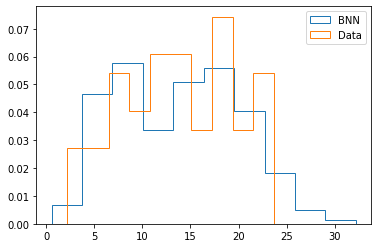}
	\caption{2020}
	\label{fig:2020}
\end{subfigure}
\begin{subfigure}{0.2\textwidth}
	\centering
	\includegraphics[width=0.9\textwidth]{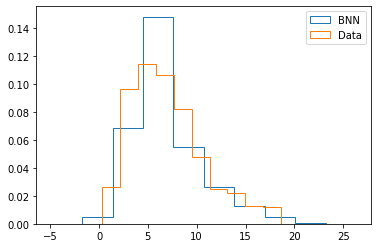}
	\caption{2021}
	\label{fig:2021}
\end{subfigure}
\caption{Comparing the annual wind speed distribution predicted by the BNN with the real data. Results are shown for the years after the wind farm was constructed.}\label{wind_speed_dist}
\end{figure}

When designing a new wind farm, practitioners are not interested in good predictions of the hourly wind speed but more interested in good predictions of the wind speed distribution over long periods of time. Figure \ref{wind_speed_dist} shows that the BNN predicted distributions, post wind-farm construction, agree well with the real data distributions and therefore show the value that our wind speed BNN can bring to wind farm planning.

\section{Conclusion}\label{sec:conc}
In this work, we have shown that neural networks can be used to predict both wind direction and wind speed from ocean data. We have also shown the value that can be added to results through the use of Bayesian neural networks (BNNs), which provide the probability of each outcome and give a measure of the uncertainty of the model and data. In particular, BNNs allow the user to see whether the accuracy and uncertainty of the wind speed and direction predictions is affected by the construction of a wind farm or not. If unaffected, as for the site considered in this work, this means that pre-farm data can be used to make wind-field estimates for after the farm's construction, without the need for further training. In further work, we plan to use this capability to assist in the design optimisation of future offshore wind farms.

\section*{Acknowledgements}
This work was supported by Towards Turing 2.0 under the EPSRC Grant EP/W037211/1  and The Alan Turing Institute. Data was made available by the FINO (Forschungsplattformen in Nord- und Ostsee) initiative, which was funded by the German Federal Ministry of Economic Affairs and Climate Action (BMWK) on the basis of a decision by the German Bundestag, organised by the Projekttraeger Juelich (PTJ) and coordinated by the German Federal Maritime and Hydrographic Agency (BSH).

\bibliographystyle{chicaco}
\bibliography{references.bib}

\end{document}